\documentclass[10pt,twocolumn]{IEEEtran}
\usepackage{graphicx}          
\usepackage{amsmath}
\usepackage{amsfonts}
\usepackage{url}
\usepackage{times,epsfig}
\usepackage{psfrag}
\usepackage{subfigure}
\usepackage{stfloats}
\usepackage{amssymb}
\usepackage{multirow}
\usepackage{colortbl}
\usepackage[justification=centering]{caption}
\usepackage{fancyhdr}
\begin{document}

\pagenumbering{arabic}

\title{Reliable Federated Learning for Mobile Networks}

\author{Jiawen Kang, Zehui Xiong, Dusit Niyato, \emph{Fellow, IEEE}, Yuze Zou, Yang Zhang,  Mohsen Guizani, \emph{Fellow, IEEE}
	
\IEEEcompsocitemizethanks{
Jiawen Kang, Zehui Xiong, Dusit Niyato are with School of Computer Science and Engineering, Nanyang Technological University, Singapore. 
Yang Zhang is with School of Computer Science and Technology, Wuhan University of Technology, China, and also with School of Computer Science and Engineering, Nanyang Technological University, Singapore. 
Yuze Zou is with School of Electronic Information and Communications, Huazhong University of Science and Technology, China. 
Mohsen Guizani is with the Department of Computer Science and Engineering, College of Engineering, Qatar University, Qatar. (\textit{Corresponding author: Zehui Xiong})
}
}
\maketitle
\pagestyle{headings}

\begin{abstract}
%Federated learning as a promising machine learning approach has emerged that leverages distributed personalized dataset from a number of nodes, e.g., mobile devices, to improve performance while simultaneously providing privacy preservation for mobile users. In the federated learning, training data is widely distributed and maintained on the mobile devices. A central aggregator updates a global model by collecting local updates from mobile devices using their local training data to train the global model in each iteration. However, data owners (i.e., the mobile devices) acting as workers may generate unreliable local updates to mislead federated learning tasks. The workers may perform the unreliable updates intentionally, e.g., the data poisoning attack, or unintentionally, e.g., low-quality data caused by energy constraints or high-speed mobility. Therefore, it is crucial to select reliable or trusted workers for the federated learning tasks. In this article, we first introduce the reputation as a metric and propose a reputation-based worker selection scheme for reliable  federated learning. To achieve efficient reputation management without repudiation and tampering, we leverage consortium blockchain as the enabling technology to manage the reputation of the workers in a decentralized way. Numerical results demonstrate that the proposed schemes are reliable for federated learning in the mobile networks.

Federated learning, as a promising machine learning approach, has emerged to leverage a distributed personalized dataset from a number of nodes, e.g., mobile devices, to improve performance while simultaneously providing privacy preservation for mobile users. In the federated learning, training data is widely distributed and maintained on the mobile devices as workers. A central aggregator updates a global model by collecting local updates from mobile devices using their local training data to train the global model in each iteration. However, unreliable data may be uploaded by the mobile devices (i.e., workers), leading to frauds in tasks of federated learning. The workers may perform  unreliable updates intentionally, e.g., the data poisoning attack, or unintentionally, e.g., low-quality data caused by energy constraints or high-speed mobility. Therefore, finding out trusted and reliable workers in federated learning tasks becomes critical. In this article, the concept of reputation is introduced as a metric. Based on this metric, a reliable worker selection scheme is proposed for federated learning tasks. Consortium blockchain is leveraged as a decentralized approach for achieving efficient reputation management of the workers without repudiation and tampering. By numerical analysis, the proposed approach is demonstrated to improve the reliability of federated learning tasks in mobile networks.
\end{abstract}

\begin{IEEEkeywords}
Federated learning, consortium blockchain, reputation management,   mobile networks.
\end{IEEEkeywords}

%%%%%%%%%%%%%%%%%%%%%%%%%%%%%%%%%%%%
\section{Introduction}

Mobile devices, such as smart phones or vehicles, equipped with a variety of sensors generate a huge amount and diverse types of user data \cite{kang2018towards}. Recently, for greatly improving mobile services and enabling smarter mobile applications, it is increasingly popular to utilize machine learning technologies to train models on such user data, e.g., service recommendation and mobile healthcare \cite{cui2018survey}. However, a majority of machine learning technologies require a large amount of user data with sensitive privacy information to be aggregated in a central server for model training and analysis. This results in exorbitant  communication and storage cost, and the mobile users are at risk of serious privacy leakage \cite{dibconsortium}.

To address the privacy challenges, a decentralized machine learning paradigm called federated learning has been proposed to enable mobile devices (e.g., vehicles) to collaboratively train a global model required by a central aggregator (i.e., a task publisher) in a decentralized manner, without the need of centrally storing raw training data. In the federated learning, the mobile devices download a global model from the central aggregator in each iteration, and then train and improve the current global model by using their local raw data. The mobile devices send the local model updates to the  central aggregator.  By  aggregating these local model updates, the central aggregator generates a new global model for the next iteration.  Both the mobile devices and the central aggregator repeat the above process until the global model achieves a certain accuracy \cite{anh2018efficient}. This paradigm significantly reduces risks of sensitive privacy leakage by decoupling of model training from the need for direct access to the raw training data \cite{dibconsortium}.

Although federated learning brings great benefits for mobile networks, it is still susceptible to various adversarial attacks  in its primary stage. That is, during a federated learning process, data owners  may mislead a global model by intentional or unintentional behaviours \cite{shayan2018biscotti}. For intentional behaviors, an attacker can send malicious updates, i.e., the poisoning attack, to affect the global model parameters resulting in the failure of current collaborative learning.  The authors in \cite{fung2018mitigating} demonstrated the vulnerability of federated learning to sybil-based poisoning through experiments, and showed that existing defenses to such attacks are ineffective.
In addition, much more dynamic mobile networking environments indirectly result in some unintentional behaviors of data owners. The data owners may also indeliberately update low-quality models caused by high-speed mobility or energy limitation, thus adversely affecting the federated learning. Therefore, it is of paramount importance for federated learning to defend against such intentionally and unintentionally unreliable  local model updates.

In this article, we propose that reputation can be used to provide solutions to select reliable and trusted workers for the federated learning tasks. Existing studies show that reputation can reflect the rating of how reliable or trusted an entity is in certain activities according to its historical  behaviors \cite{kang2018towards,iotj2018}. Along with this direction, we are motivated to treat the reputation as a fair metric and design a reputation-based worker (i.e., data owner) selection scheme for reliable federated learning. With the help of reputation, each task publisher selects only high-reputation workers to eliminate the impact from unreliable workers, thereby leading to high accuracy of the learning task \cite{shayan2018biscotti}. Each task publisher calculates reputation opinions of every interacting worker through a subjective logic model. In the subjective logic model, the task publishers integrate their own opinions based on past interactions and recommended opinions from other task publishers \cite{kang2018towards,iotj2018}. All the reputation opinions of the task publishers for the workers should be recorded in a non-repudiation and tamper-resistance manner for reliable reputation calculation.

To realize reliable reputation calculation as well as reputation management in federated learning, we design a consortium blockchain acting as a trusted and decentralized ledger to record and manage the data owners' reputation. The consortium blockchains are specific blockchains that perform the consensus process on pre-selected miners with mild cost in a short time  \cite{kang2018towards,iotj2018}.  In the mobile networks, edge nodes, e.g., roadside units and based stations, are commonly deployed over the networks and easily reachable by task publishers and mobile devices, can be the pre-selected miners owing to having sufficient storage and computation resources \cite{iotj2018}. The reputation values of the data owners are securely managed and stored on the consortium blockchain consisting of the edge nodes. The consortium blockchain is an efficient and practical blockchain technology running light-weight and fast consensus mechanisms on the miners.
The major contributions of this article are summarized as follows:
\begin{itemize}
\item To defend against unreliable model updates, reputation is introduced as a reliable metric to select trusted   workers for reliable federated learning.
\item A multi-weight subjective logic model is applied to design an efficient reputation calculation scheme according to both task publishers' interaction histories and recommended reputation opinions.
\item To achieve secure reputation management, the reputation is managed in a decentralized manner by employing  the consortium blockchain deployed at edge nodes.
\end{itemize}

\section{Federated Learning and Its Vulnerabilities}

\subsection{Federated Learning and Its Mobile Applications}
Traditional machine learning methods train models by using   training data stored in a centralized server or dataset. But these methods face several critical challenges including single point of failure, sensitive data leakage, and huge overhead to collect and store the training data. To overcome these challenges, Google introduced a promising technique named federated learning that allows distributed mobile devices to collectively train a global model using their raw data while keeping these data locally stored on the mobile devices. Every mobile device computes a local update, for example via a distributed Stochastic Gradient Descent (SGD) algorithm,  and uploads the local update, i.e., weight parameters of current global model,  to a central aggregator. The central aggregator, e.g., a central server, collects all the local updates and calculates the average value of these local updates as a new  global model. The federated learning significantly improves privacy protection of the mobile devices through blocking attack surfaces for direct access to the raw training data \cite{dibconsortium}.

With the increasing popularity of federated learning, more and more mobile applications with the federated learning  have emerged. Some typical applications  are listed as follows.
\begin{itemize}
	\item \emph{Google keyboard:} Gboard\footnote{https://ai.googleblog.com/2017/04/federated-learning-collaborative.html}   as a virtual keyboard application from Google employs federated learning to improve language model quality, while simultaneously offering  security and privacy protection for users by training input data locally. 
	
   \item \emph{Service recommendation:} 
   Service providers collect searching and location histories from  mobile devices to train  entertainment and restaurant recommendation systems for enhancing service quality, but may cause serious privacy leakage risks for the mobile users. To ensure privacy preservation, the mobile devices join  training recommendation models without concerning about privacy leakage by using the federated learning.

   \item \emph{Traffic monitoring and prediction:} 
   UberEATs\footnote{https://eng.uber.com/michelangelo/} leverages real-time traffic information to calculate estimated time of food delivery in a distributed learning manner. However, the distributed vehicles  are not willing to share  local traffic sensing data  because of the concern of  privacy leakage. To address this problem, the federated learning technique can be used to train prediction models without a direct access to the personal  data on the vehicles, which not only enhances traffic prediction accuracy  but also  protects data privacy of vehicles \cite{liang2018towards}.

   \item  \emph{Mobile healthcare:} Health data from patients can be shared among hospitals or medical researchers to improve clinical services and healthcare analytics. Sharing such data with sensitive privacy information is facing serious challenges in mobile healthcare. Federated learning  therefore is introduced to avoid centrally health data collection and collaboratively train models by using local health data in the mobile devices.  NVIDIA Clara\footnote{https://blogs.nvidia.com/blog/2018/10/10/kings-college-london-nvidia-clara/}  is used to deploy federated learning tasks to recommend the best treatment or automatic biomarker determination.
\end{itemize}

\subsection{Security Challenges and Motivations}
Although the federated learning is promising to be applied in mobile environments, some critical challenges exist including  reliable and trusted worker selection problems for model training.  On the one hand, due to the openness and complexity of mobile network architectures, the data owners performing maliciously unreliable updates may result from: (i) sensing data from malicious intent or tampered devices may include deceptive information, which is similar to false data injection attacks in smart grids \cite{sg2}; (ii)  the data can be arbitrarily manipulated when being transmitted through insecure communication channels \cite{dibconsortium,sg2}. If a malicious data owner is selected to be a worker, the malicious worker may intentionally launch or collude with other workers to launch attacks, such as \emph{poisoning attacks}~\cite{shayan2018biscotti}. For the poisoning attacks,  malicious workers deliberately tamper with a fraction of training data or  inject  poisonous data into the training datasets to increase the probability of misclassification thus  manipulating the results of  training models \cite{iotj2018}.  
On the other hand, the data owners may inadvertently provide unreliable local update from low-quality raw data because of energy constraints or high-speed mobility. Both the intentional and unintentional behaviors can degrade the quality of  the  global model managed by a central aggregator\footnote{We assume that the central aggregator is not compromised or malicious.}, hence affecting the final  outputs of the global  model \cite{shayan2018biscotti}.  Therefore, it is vitally important to design a reliable  worker selection scheme during model training.  Nevertheless, in the federated learning, the following challenges for the worker selection need to be addressed.
\begin{itemize}
\item \emph{No reliable and fair metrics to evaluate workers}: A majority of federated learning systems randomly select  mobile devices to be the  workers through verifiable random functions \cite{shayan2018biscotti} or resource conditions \cite{anh2018efficient}. However, the existing schemes cannot measure the trustworthiness level of workers to remove unreliable or untrusted  workers.

\item \emph{No efficient and universal worker selection schemes}: The  workers in  existing federated learning schemes  are selected either by a centralized authority, or by a decentralized method that all mobile devices  join model training at will. As a result, the worker selection schemes are suffering  from negative influence of unreliable or untrusted workers. For the federated learning in mobile networks,  it is difficult to design an efficient and universal worker selection scheme for identifying high-quality data contributors and malicious worker candidates.

\item \emph{No  timely monitoring methods for workers}: It is hard for the central aggregator/server (i.e., task publisher) to  monitor the large-scale worker behaviors in real-time. The central aggregator  without timely and dynamic monitoring methods cannot detect and remove the malicious or unreliable workers from the system. As a result, a malicious or unreliable worker may be selected to be a worker again for a new federated learning task  because of the lack of time-accumulated metrics to evaluate the  worker's historical performance and the synchronous information of  malicious and unreliable worker lists.
\end{itemize}
To cope with the  above challenges, we introduce a reliable metric and design a reliable worker selection scheme for federated learning in mobile networks.

\section{Reputation Management for Reliable Federated Learning}

%%%%%%%%%%%%%%%%%%%%%%%%%%%%%%%%%%%%%%%%%%%%%%%%%%%%%%%%%%%%%%%%%%%%%%%%%
\begin{figure*}[t]\centering
  \includegraphics[width=16.5cm,height=7cm]{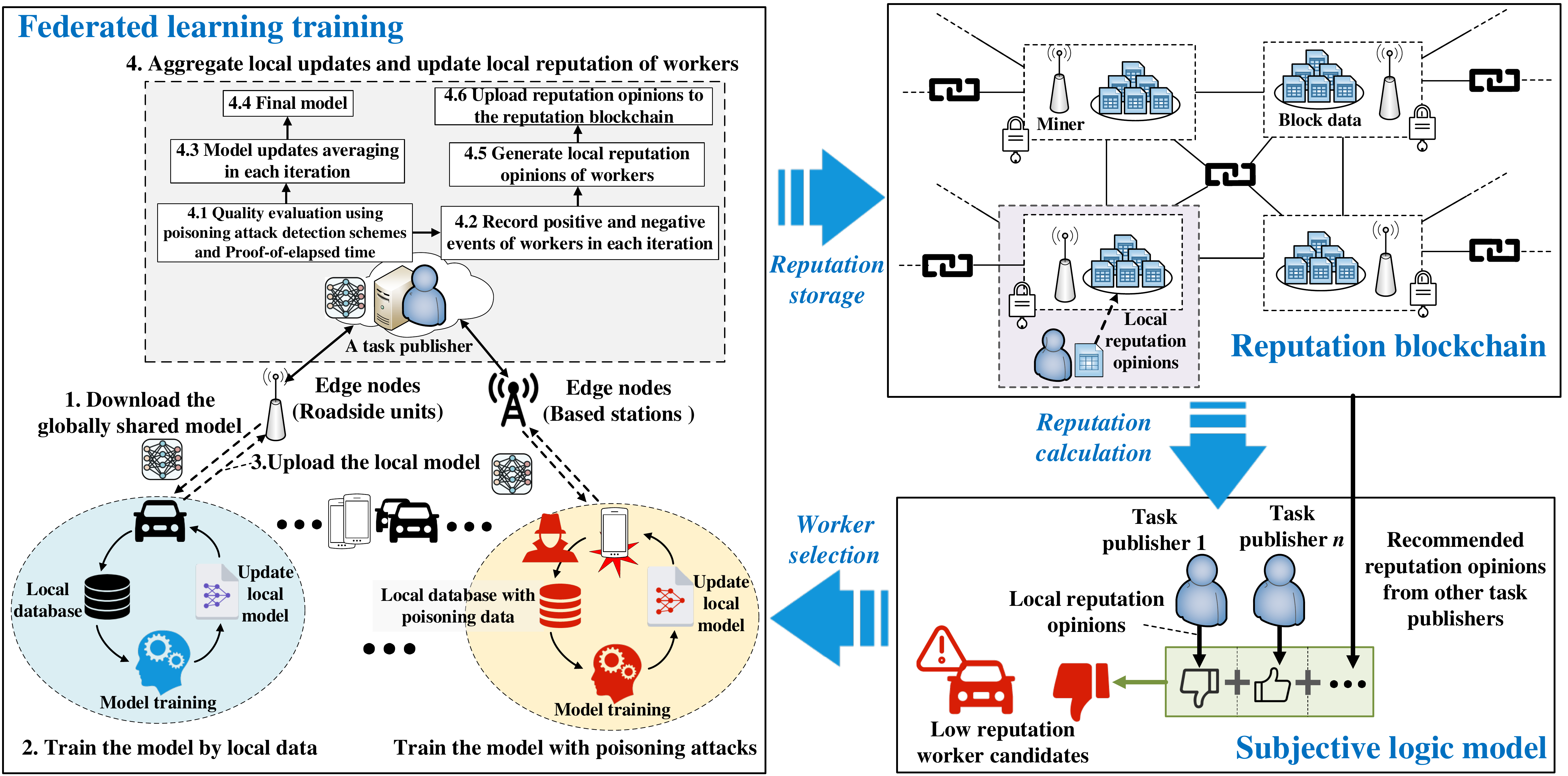}
  \caption{ Consortium blockchain-based reputation management for secure federated learning.}
   \label{systemmodel}
 \vspace*{-4mm}
\end{figure*}
%%%%%%%%%%%%%%%%%%%%%%%%%%%%%%%%%%%%%%%%%%%%%%%%%%%%%%%%%%%%%%

\subsection{Overview of Reputation Management in Crowdsensing}
Regarding data quality problems, recent studies mainly focus on introducing reputation as a metric to identify  whether a data provider is honest or malicious and evaluate its data quality \cite{kang2018towards,iotj2018}, especially in crowdsensing scenarios \cite{twoconsensus,2017quantifying,xie2015incentive}.  The reputation is used to select data providers who are more likely to provide high-quality data in crowdsensing.
An~\emph{et~al.}~\cite{twoconsensus}  proposed a data provider selection scheme by using credit matching degree and trajectory matching degree for improving data quality in crowdsensing. The credit matching degree is calculated to measure the possibility that the worker submits high-quality data.  Xie~\emph{et~al.}~\cite{xie2015incentive}  designed a reputation mechanism to prevent low-skilled workers and encourage high-skilled workers to participate in the crowdsensing tasks. The reputation values of workers are obtained and updated according to their historical contributions. Pouryazdan~\emph{et~al.}~\cite{2017quantifying} proposed a collaborative reputation scoring method based on statistical and vote-based user reputation scores to quantify the data trustworthiness, which  improves platform utility and data trustworthiness in mobile crowdsensing.

Inspired by the great potential of reputation in solving data quality problems in crowdsensing, we  adopt the reputation metric to the selection of  trusted and reliable workers for enhancing model training performance in the federated learning. The reputation can reflect  how well a worker has performed about model training, which can be measured from its training  task completion history with the past behaviors of good or unreliable activities \cite{twoconsensus}.  { With the help of reputation, task publishers select trusted and reliable workers to train the  global model well,  which can prevent the poisoning attacks launched by malicious workers and also remove unreliable data providers for obtaining high accuracy of the global model. The recent studies on federated learning \cite{shayan2018biscotti,kim2018device} have indicated that a central aggregator is vulnerable to security problems,   e.g., single point of failure \cite{twoconsensus}.
In this article,  to avoid the potential risks of central reputation calculation and management,  we employ a decentralized reputation calculation method named subjective logic model \cite{iotj2018}, and a consortium blockchain with the properties of immutability and decentralization to realize secure  reputation management  \cite{kang2018towards}.  Compared with centralized reputation management, consortium blockchain as a decentralized ledger can manage the reputation in a real-time and parallel  manner without large computation overload. Similar to \cite{iotj2018}, the consortium blockchain performs the consensus process on pre-selected miners with mild cost in a short time, which is particularly suitable and practical for mobile networks because of light-weight and faster consensus agreement. More details about the reputation management and  the subjective logic model for reputation calculation  are presented in  Section III-B and Section IV, respectively.}

\subsection{Reputation-based Worker Selection Scheme with Consortium Blockchain}

As shown in Fig.~\ref{systemmodel}, the mobile devices collect local sensing data and generate various user data from mobile applications. Mobile applications with federated learning perform model training by using these data without the need of data aggregation for privacy preservation. The detailed steps about the federated learning are shown as follows \cite{iotj2018}.

\begin{itemize}
\item \textbf{Step 1:}  \textit{Task publishment}:
Federated learning tasks from task publishers are first broadcast with specific  data requirements (e.g., data sizes,  types and time range). Mobile devices, that want to join one task and also satisfy the specific data requirements, will send a joining request with identity and data resource information back to one task publisher.

\item \textbf{Step 2:}  \textit{Worker selection}: The task publisher validates the identity and data resource information of the requesters, then the legal requesters can be the worker candidates. The task publisher starts to select its  workers from the worker candidates according to their reputation values calculated by the subjective logic model in Section IV. The worker candidates with reputation values above a threshold can be selected as the workers. Here, the task publishers can set different reputation thresholds by themselves according to their own security level requirements. Without loss of generality, all task publishers can choose the same reputation threshold for current federated learning tasks. The reputation thresholds can also be adjusted by some statistical metrics based on the mean and standard deviation of reputation values of their worker candidates. The reputation values of the worker candidates are calculated according to (i)  local reputation opinions generated from direct interaction histories, and (ii) recommended reputation opinions of other task publishers stored on an open-access consortium blockchain named reputation blockchain. The reputation blockchain with decentralization and tramper-resistant natures is a public ledger established on the pre-selected miners, which records the reputation opinions into data blocks. These reputation opinions in the data blocks are  transparent and tramper-proof evidence even if  damage occurs  \cite{kang2018towards,iotj2018}. 

\item \textbf{Step 3:} \textit{Reputation calculation}: The task publisher utilizes the subjective logic model to generate local reputation opinions for the worker candidates based on interaction histories. The subjective logic model takes three weights about the past interactions into consideration to form the local opinions for each worker candidate. By combining the
local reputation opinions with recommended reputation opinions, the task publishers generate  composite  reputation as the final reputation  for each worker candidate. The recommended reputation opinions can be downloaded from the reputation blockchain and obtained from the latest block data.  More details about reputation calculation are  depicted  in Section  IV.

\item \textbf{Step 4:} \textit{Federated learning}: We can adopt different optimization algorithms to train a federated learning model. In this article, we  utilize SGD algorithm\footnote{ {For obtaining $\varepsilon$-accuracy, the SGD algorithm needs $O(\mu^2/\varepsilon)$ iterations on each worker, where $\mu$ is the  condition number defined as the ratio of smoothness and strong convexity parameters of a strongly convex problem \textit{P}.}} that iteratively selects a batch of training examples to calculate their gradients against the current model parameters and takes gradient steps in the direction that minimizes the loss function  \cite{shayan2018biscotti}. 
The task publisher first randomly chooses  an initial SGD model (i.e., initial parameters) from predefined ranges as the  global model. This initial SGD model is received by selected workers and the workers collaboratively train the global model by using their own local data. The workers generate local model updates and the corresponding local computation time and upload these information to the task publisher. The local computation time is used to verify the reliability and authenticity of local model updates by comparing the data size of training data, in which the local computation time is proportional to  the data size. To ensure the truthfulness  of local computation time, we consider employing the proof of elapsed time method under the Intel's SGX technology  \cite{kim2018device}. After validating the computation time, the task publisher can determine the ``lazy" workers that have not trained all of the local  data.

Moreover,  some poisoning attack detection schemes are carried out by the task publisher to identify the poisoning attacks and unreliable workers.  Typical detection schemes include Reject on Negative Influence (RONI) scheme \cite{shayan2018biscotti}    for Independent and Identically Distributed (IID) scenarios  and FoolsGold scheme  \cite{fung2018mitigating}  for non-IID scenarios.
With the help of these schemes,  the task publisher removes  malicious updates from poisoning attacks and  unreliable local model updates from the lazy or untrusted workers. Then, the task publisher generates a new global model by calculating the average of  the rest of local model updates. 
Similar to  \cite{shayan2018biscotti}, we consider that the distribution of data among mobile devices is sufficiently uniform to enable  the RONI work well in gradient validation.
The task publisher sends the new global model to the selected  workers for the next model iteration until  the  global model meets the predefined convergence conditions. The workers obtain rewards from the task publisher according to their data contribution and model training behaviors for the federated learning task \cite{shayan2018biscotti,kim2018device}.   During the federated learning process, either the lazy and unreliable workers or the workers with poisoning attackers in each model iteration  are recorded as a negative interaction  by the task publisher. Finally, the task publisher generates local reputation opinions for the workers based on their performance  in the federated learning task.

\item \textbf{Step 5:} \textit{Reputation updating}:  To achieve secure reputation management, the task publisher's  interaction histories and  local reputation opinions  for the workers with digital signatures  are recorded as ``transactions" and uploaded to the pre-selected miners in the reputation blockchain. These  miners  execute consensus algorithms, such as PBFT, and the reputation opinions and interaction histories are stored as a data block to be added into the reputation blockchain. After that, all task publishers can obtain the latest reputation opinions for a certain worker candidate from the reputation blockchain.  Lastly, with the help of the reputation blockchain,  the task publishers are able to select high-reputation workers for federated learning tasks.
\end{itemize}

\section{Efficient Reputation Calculation Scheme}
To assess the trustworthiness  of a worker candidate, reputation opinions from task publishers should be collected and integrated into  a composite reputation value of the worker candidate for secure worker selection. We therefore utilize the subjective logic model to calculate  composite reputation values of worker candidates. Subjective logic is widely used to evaluate the trust  level between different entities in the networks \cite{kang2018towards,iotj2018}, which is a specific framework of uncertain reasoning that uses a belief metric named ``opinion" to represent a subjective beliefs about the world.  The opinion is denoted by a tuple consisted of belief, distrust, and uncertainty to express the subjective belief of an entity or an event.  For example, in vehicular networks, a task publisher performs a federated learning-based traffic prediction service with the help of vehicles. The task publisher's subjective belief for a vehicle increases if the publisher believes that the  model updates provided by  the vehicle are high-quality without the external impacts of unstable communication link between them, and vice versa.
All reputation opinions from task publishers are securely updated and stored in the decentralized reputation blockchain. Every task publisher selects workers by calculating  composite reputation values according to its local reputation opinion and recommend reputation opinions. More details about reputation calculation are given as follows.

\subsection{Subjective Logic Model for  Reputation Calculation }
 {During a federated learning task, e.g., vehicular service recommendation, a task publisher interacts with different vehicles (i.e., workers) for training model  corporately in each training  iteration.  By using poisoning attack detection schemes and  the proof of elapsed time scheme (Step 4 in Section III-B), the task publisher \textit{i} treats a training  iteration  as a positive interaction event   if the publisher  perceives that the local model update from a worker \textit{j} is reliable, and vice versa.  The task publisher records the numbers of positive and negative interaction events of all  workers after a learning task, i.e.,  $\alpha_j$ and  $\beta_j$,  and generates local reputation opinions for the workers. Each local reputation opinion is formally denoted as a local opinion vector consisting of i) belief degree  $b_{i \to j}$, ii) distrust degree  $d_{i \to j}$, and iii) uncertainty degree  $u_{i \to j}$.  The sum of these degrees is one.}

 {Similar to \cite{kang2018towards,iotj2018}, the uncertainty degree is determined by the quality of communication link between the worker \textit{j} and the task publisher \textit{i}, i.e., the unsuccessful probability of data packet transmission (e.g., a worker unintentionally ignoring or dropping communication packets). The belief (distrust) degree is expressed by the positive (negative) interaction percentage of  all interactions with good communication quality, denoted as ${b_{i \to j}} = (1 - {u_{i \to j}})\frac{\alpha_j }{{\alpha_j  + \beta_j }}$ and  ${d_{i \to j}} = (1 - {u_{i \to j}})\frac{\beta_j }{{\alpha_j  + \beta_j }}$.  From the local opinion vector, a local reputation value is generated to represent the task publisher's expected belief that the worker provides high-quality local model updates during the federated learning. The local reputation value is expressed as ${T_{i \to j}} = {b_{i \to j}} + \gamma {u_{i \to j}}$, where $\gamma$ is the given constant indicating an effect level of the uncertainty for the reputation.}

%%%%%%%%%%%%%%%%%%%%%%%%%%%%%%%%%%%%%%%%%%%%%%%%%%%%%%%%%%%%%%%%%%%%%%%
\begin{figure}[t]\centering
	\includegraphics[width=0.5\textwidth]{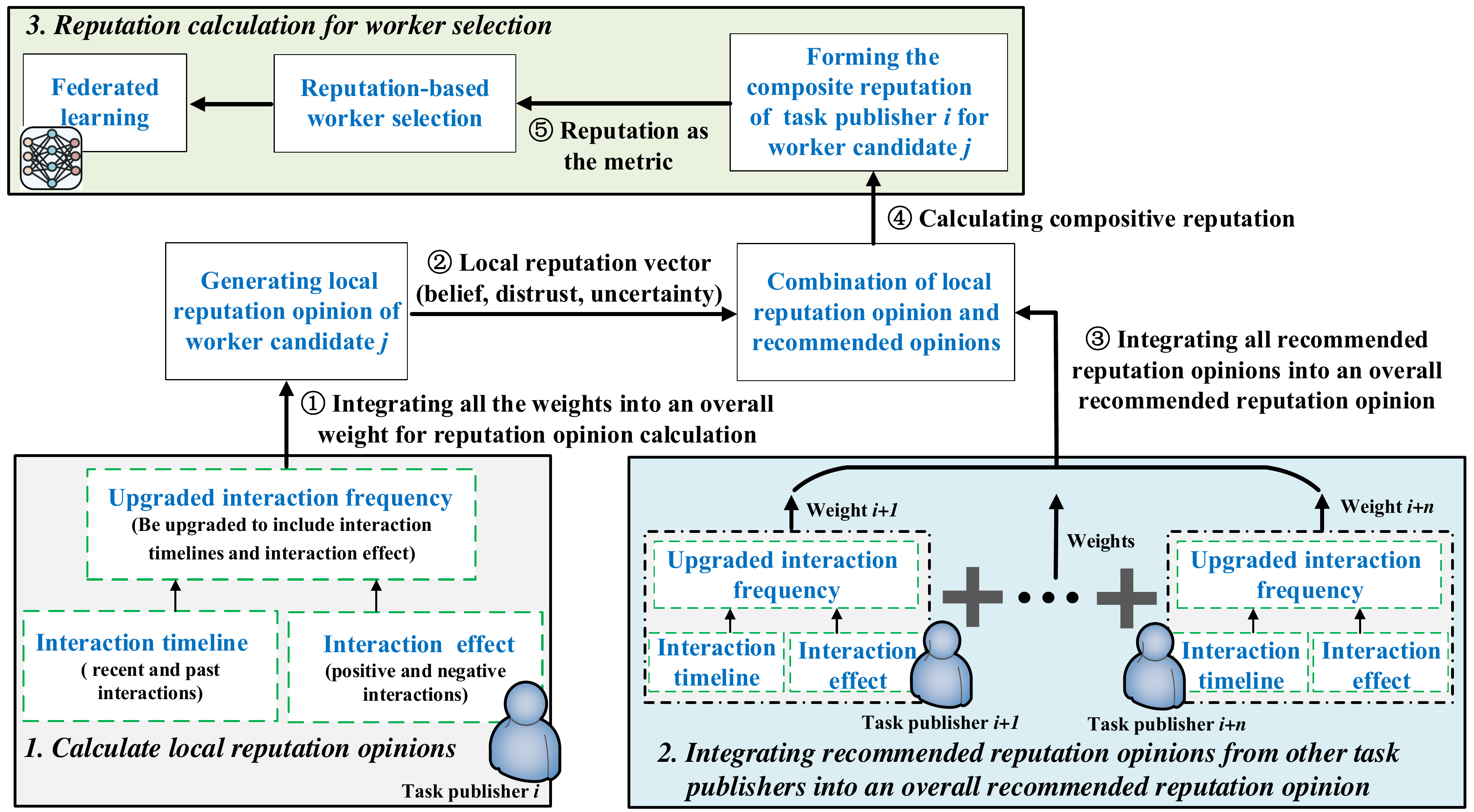}
	\caption{An overview of reputation calculation.}
	\label{ReputationCalculation}
	\vspace*{-4mm}
\end{figure}
%%%%%%%%%%%%%%%%%%%%%%%%%%%%%%%%%%%%%%%%%%%%%%%%%%%%%%%%%%%%%%

\subsection{Multi-weight Subjective Logic Model}
 Multi-weight subjective logic is an extension of subjective logic that takes different  attributes of interaction events into consideration for more accurate and reliable reputation calculation \cite{iotj2018}.  As shown in Fig.~\ref{ReputationCalculation},  we consider the following three attributes as the weights  to calculate reputation opinions.
 
\begin{itemize}
\item {\textit{Interaction Frequency:}}  The interaction frequency represents the familiarity degree between a task publisher and a worker, which is expressed by the ratio of the number of times that the  publisher interacts with the worker to the  average number of  times that the publisher interacts with other workers during a time window. The higher interaction frequency brings more prior knowledge about the worker to the publisher, hence leading to a higher local reputation opinion for the worker.

\item {\textit{Interaction Timelines:}} 
Mobile devices acting as workers are vulnerable if there is no sufficient security protection. Therefore, the workers are not always reliable or trusted in federated learning. The trust level and the local reputation opinion of a worker for the same task publisher are changing over time. To evaluate the time effects on interactions,  a time scale, e.g., three days, is utilized to  divide the interaction events into recent and past interactions. The recent interactions have a higher weight on  the task publisher's reputation opinions. 

\item {\textit{Interaction Effects:}}  Different interaction events have different effects on the reputation opinions. We classify the interaction events into: positive and negative interactions. The negative interactions, e.g., the interactions with   malicious workers or ``lazy''  workers  (judged by Step 4 in Section III-B), decrease the reputation of the workers, and vice versa. The positive interactions have a higher weight on the reputation opinion calculation.
\end{itemize}

  {Taking the interaction timelines and interaction effects into consideration, the interaction frequency is upgraded to contain the above two weights. Therefore, the interaction frequency is determined by both the two weights and the average number of  times of interactions with other workers during a time window. After that, the upgraded interaction frequency is used to generate an overall weight for local and recommended reputation opinion calculation (as shown in \raisebox{.5pt}{\textcircled{\raisebox{-.9pt} {1}}}  and \raisebox{.5pt}{\textcircled{\raisebox{-.9pt} {2}}}) \cite{iotj2018}.}
  
\subsection{Recommended Reputation Opinions}
For a task publisher, the local reputation opinions from other task publishers are treated as recommended reputation opinions. These opinions  are integrated into an overall  recommended opinion according to the task publisher's weights for each recommended opinion (as shown in \raisebox{.5pt}{\textcircled{\raisebox{-.9pt} {3}}}). The overall recommended opinion is also denoted as a recommended belief degree, a recommended distrust degree, and a recommended uncertainty degree. These degrees are calculated by weighted arithmetic mean of the belief degrees, distrust degrees and uncertainty degrees from other task publishers, respectively.

\subsection{Combining Local Reputation Opinions with Recommended Reputation Opinions}
When calculating the composite reputation value of  a worker, the task publisher takes not only the overall recommended opinions, but also its own local reputation opinion into consideration to avoid collusion cheating from other task publishers (as shown in \raisebox{.5pt}{\textcircled{\raisebox{-.9pt} {4}}}). The composite reputation of the task publisher to the worker is denoted as a final reputation opinion vector including three elements: the final belief degree, the final distrust degree, and the final uncertainty degree. The composite reputation value is determined by the final belief degree and the final uncertainty degree.  {More details about the  reputation calculation can be found in \cite{kang2018towards,iotj2018}}. With the help of the reputation metric, high-reputation worker candidates can be selected as the worker for the federated learning tasks (as shown in \raisebox{.5pt}{\textcircled{\raisebox{-.9pt} {5}}}). These high-reputation workers will train local model honestly and maintain good behaviors in the federated learning tasks for earning more profits from the system. Therefore, the reputation-based worker selection scheme can defend against unreliable local model update from intentional or unintentional data providers, hence ensuring reliable federated learning in mobile networks.

\section{Numerical Results}
\subsection{Simulation Setting}
In order to evaluate performance of the proposed schemes, we perform simulation on a well-known digit classification dataset named MNIST by using Tensorflow 1.12.0 for a digit classification. This dataset consists of  60,000 training  examples and 10,000 test examples \cite{fung2018mitigating}.  We consider ten workers in this federated learning task including two malicious workers who launch poisoning attacks, four unreliable workers with low-quality data, and four well-behaved workers.
The training sets of the well-behaved workers are randomly assigned but follows a uniform distribution over 10 classes. The  data in each unreliable worker is only assigned a certain number of classes  randomly. We employ the Earth Mover's Distance (EMD) as a metric to measure training data quality of the unreliable workers. Here, the EMD is expressed by the probability distance for a worker's data distribution compared with the actual distribution for the whole population \cite{zhao2018federated}. For the malicious workers launching poisoning attacks, they randomly receive training data with 10 classes. However, the labels of some training examples are intentionally modified for misleading  training. 
The percentage  of the modified training examples is used to indicate the attack strength.  The workers use a batch of 32 randomly sampled training examples to produce a local SGD update, and  every global model is trained with 5 synchronous iterations \cite{fung2018mitigating,iotj2018}.   Without loss of generality, the computation overhead of local model training is a constant overhead on each worker in the simulation. We establish the reputation blockchain system on the  Hyperledger Fabric v1.4.0 and use the practical and efficient PBFT algorithm with mild overhead and latency as the consensus algorithm \cite{kang2018towards,iotj2018}.

%%%%%%%%%%%%%%%%%%%%%%%%%%%%%%%%%%%%%%%%%%%%%%%%%%%%%%%%%%%%%%%%%%%%%%
\begin{figure}[t]\centering
	\includegraphics[width=0.38\textwidth]{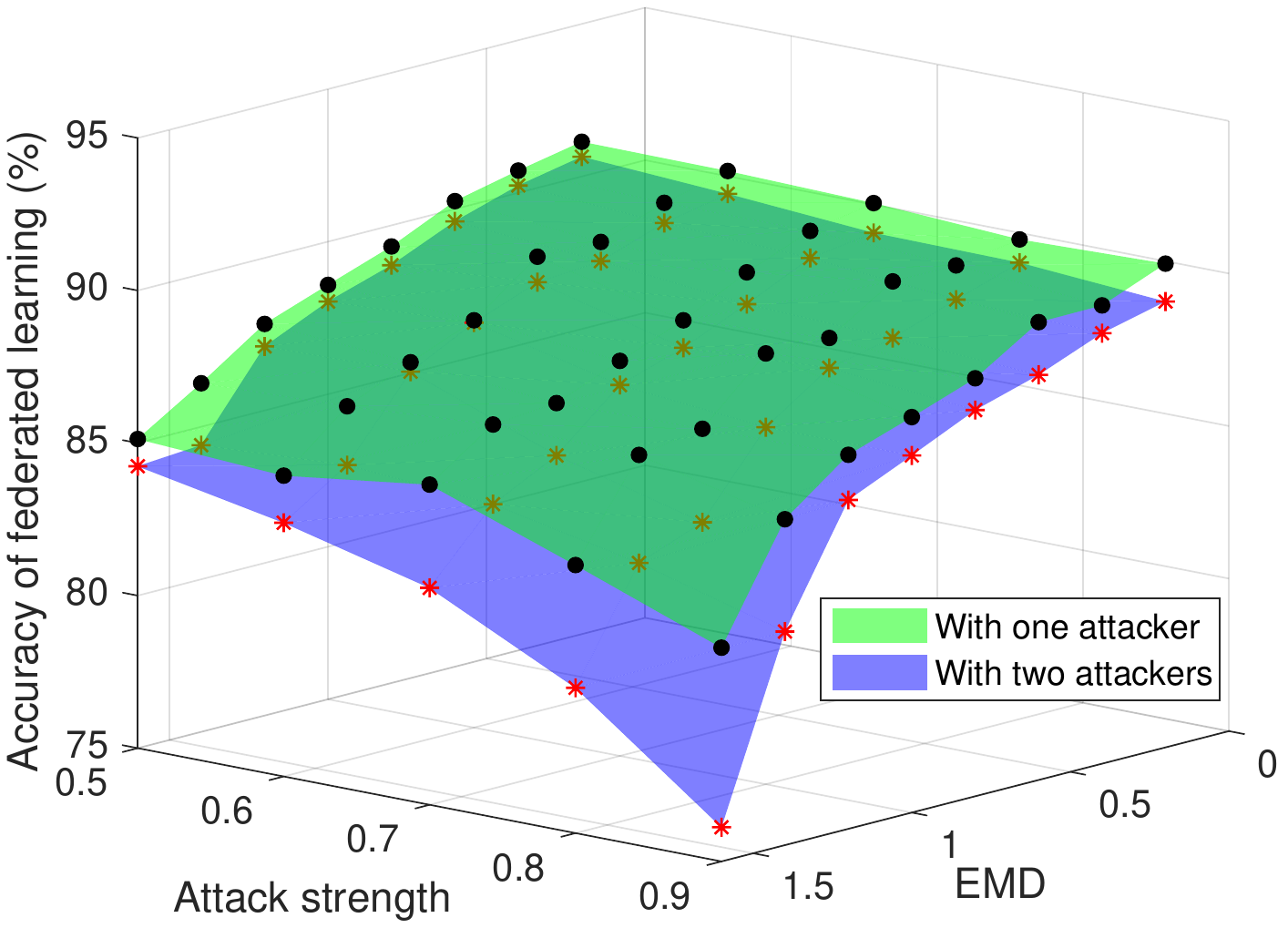}
	\caption{The accuracy comparison with respect to attack strengths and  data quality levels.}
	\label{accuracy}
	\vspace*{-4mm}
\end{figure}
%%%%%%%%%%%%%%%%%%%%%%%%%%%%%%%%%%%%%%%%%%%%%%%%%%%%%%%%%%%%%

For reputation calculation, the interaction frequency between task publishers and workers is from 20 to 40 federated learning tasks every week.  The weight parameters of negative, positive, recent, and past interactions, and the time scale in the proposed  Multi-weight Subjective Logic (MSL) scheme are referred to \cite{kang2018towards}.   The unsuccessful transmission probability of data packets ranges from 0\% to 40\%, and the initial reputation of all the workers is 0.5. 
We compare the proposed MSL scheme with a  Traditional Subjective Logic (TSL) scheme from~\cite{iotj2018}, and an Aggregated Trust Value (ATV) scheme referred to ~\cite{yang2018blockchain}. In the ATV scheme, reputation is calculated by aggregating trust value offsets with different weights from the task publishers. The trust value offset is determined by the ratio of the difference between positive events and negative events to the total number of events.

\subsection{Performance Results}

Figure~\ref{accuracy} shows the federated learning accuracy with respect to different  poisoning attack strengths and EMDs. There are three factors that affect the learning accuracy: EMD, attacker number, and attack strength. The increase of any one of the above factors leads to the decrease of accuracy. The unreliable and untrusted workers with low-quality training data have negative impacts on the accuracy. For example, as shown in Fig.~\ref{accuracy},  the learning accuracy in the case with two attackers is only 76.12\%, which is 7.7\% lower than that with one attack when EMD is 1.6 and the attack strength is 0.9.

%%%%%%%%%%%%%%%%%%%%%%%%%%%%%%%%%%%%%%%%%%%%%%%%%%%%%%%%%%%%%%%%%%%%%%
\begin{figure}[t]\centering
	\includegraphics[width=0.35\textwidth]{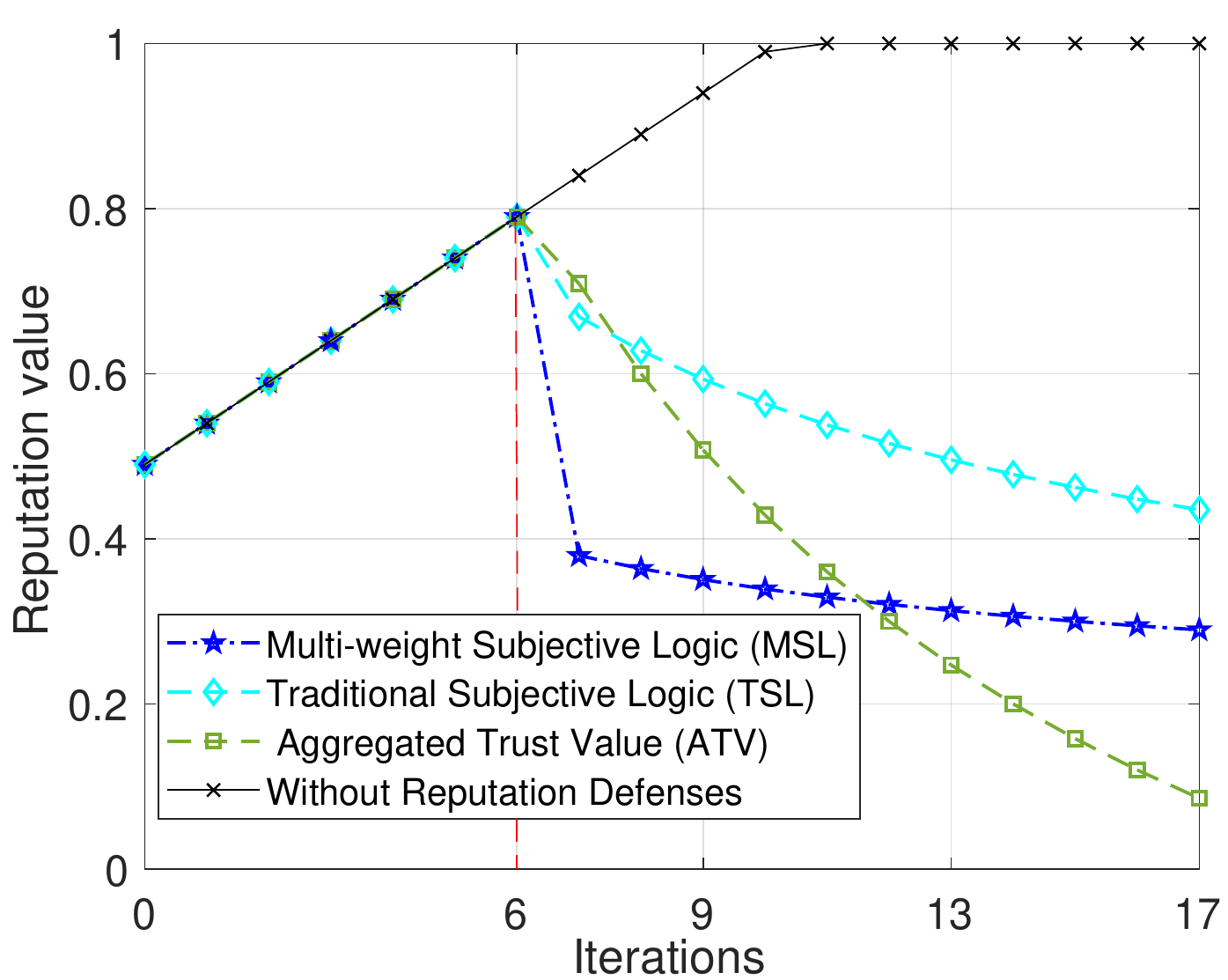}
	\caption{Reputation changes based on behavious of a worker.}
	\label{reputationchange}
\vspace*{-4mm}
\end{figure}
%%%%%%%%%%%%%%%%%%%%%%%%%%%%%%%%%%%%%%%%%%%%%%%%%%%%%%%%%%%%%

To illustrate the reputation change of a malicious or unreliable worker, we set that this worker performs well in the former 6 federated tasks on purpose to increase its reputation value. Then, the worker trains local models on its poisoning or unreliable examples for 30 task publishers with the probability of 0.8.  As shown in Fig.~\ref{reputationchange},  when the worker performs misbehaviours, its reputation begins to decrease in the MSL, TSL and ATV schemes, but  is still linearly increasing in the scheme without reputation defenses. Due to considering the interaction effects, frequency and timeline, the reputation of the MSL scheme has a  sharper and larger decrease than those in  the ATV and TSL schemes in a short time. Moreover, the reputation of the ATV scheme drops faster than that of the MSL scheme after 12 iterations because the ATV scheme merely focuses on the interaction effects when calculating trust value offsets.

\begin{figure}[t]\centering
	\includegraphics[width=0.35\textwidth]{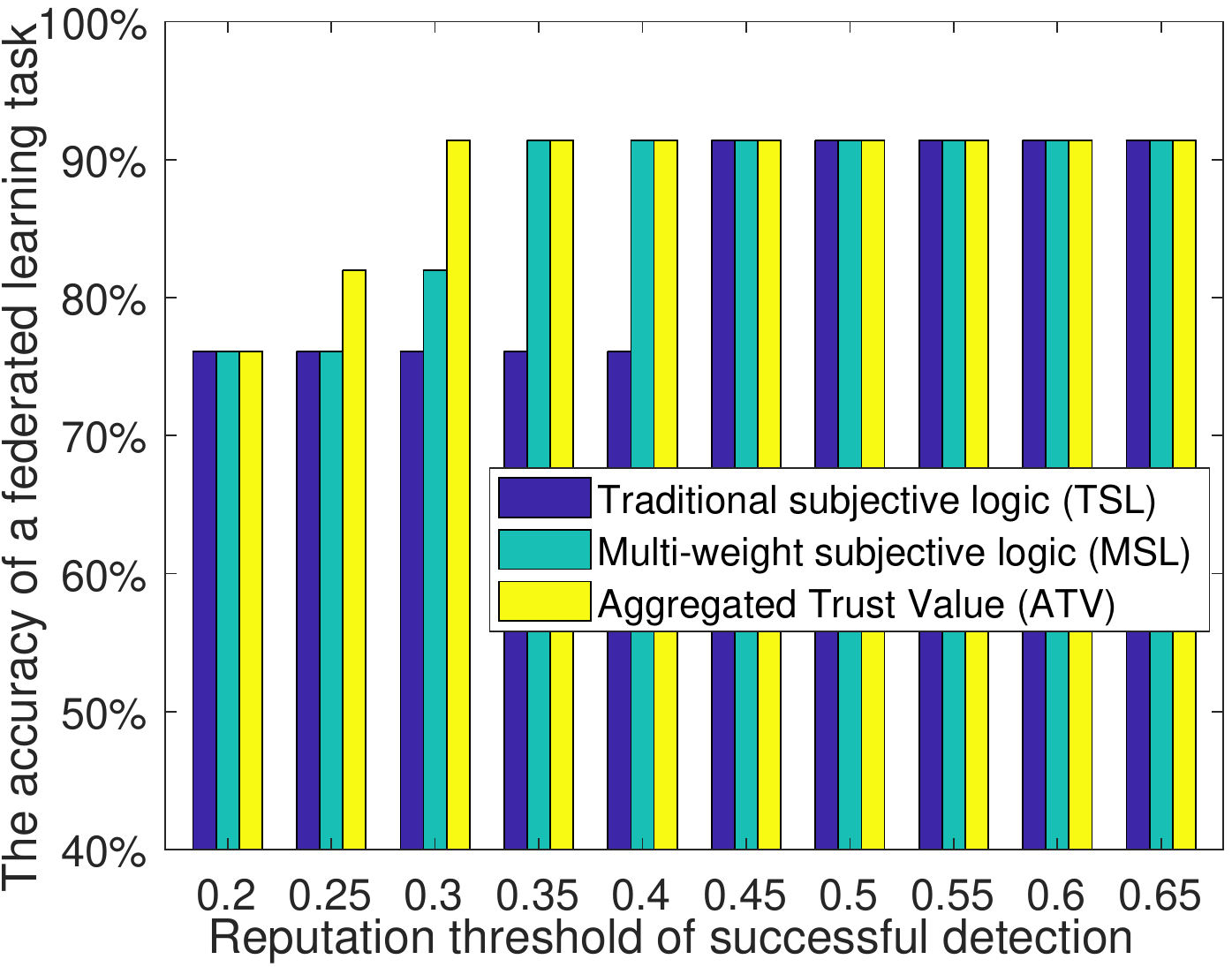}
	\caption{The impact of reputation thresholds of successful detection on the federated learning accuracy.}
	\label{FL_detectRate}
	\vspace*{-4mm}
\end{figure}

Figure~\ref{FL_detectRate} shows the impact of reputation thresholds of successful detection  on the  accuracy of a federated learning task (EMD=1.6, attack strength=0.9). If a worker's calculated  reputation is below the given reputation threshold, the worker will be treated as a malicious or unreliable worker.  Figure~\ref{FL_detectRate} illustrates that the higher reputation threshold brings a higher  federated learning accuracy. Although the accuracy of the MSL scheme is lower than that of the  ATV scheme under lower reputation thresholds, the MSL scheme has the same performance as that of the ATV scheme when the reputation is higher than 0.35. The reason is that the ATV scheme is sensitive to current negative events but ignores the well-behaved  histories for good worker candidates with unintentional mistakes. This can result in false-positive errors and partial reputation calculation to reduce the incentive of the worker candidates. The TSL, MSL and ATV schemes achieve the same performance when the reputation threshold is above 0.45. This reason is that the malicious and unreliable workers are easier to be detected and hence removed in the case of high EMD and attack strength. In summary, the MSL scheme can achieve a more accurate and fair reputation calculation, thereby leading to a more reliable worker selection in federated learning.

%%%%%%%%%%%%%%%%%%%%%%%%%%%%%%%%%%%%%%%%%%%%%%%%%%%%%%%%%%%%%%%%%%%%%%
%%%%%%%%%%%%%%%%%%%%%%%%%%%%%%%%%%%%%%%%%%%%%%%%%%%%%%%%%%%%%

\section{Conclusion and Future Directions}

In this article, we addressed worker selection issues to ensure reliable federated learning in mobile networks. A reputation-based scheme was designed to select reliable and trusted workers. For efficient and secure reputation management, we calculated workers' reputation by using  multi-weight subjective logic model, and employed  consortium blockchain to manage the reputation  with tamper resistance and non-repudiation in  a decentralized manner. Numerical results  showed that our schemes can bring reliable federated learning to the mobile networks.
There are several possible directions that are worth being studied:
I) Due to model update validation limitation of RONI scheme in non-IID setting, more accurate and efficient  validation schemes for non-IID datasets should be designed to improve the detection performance of poisoning attacks in the proposed worker selection schemes.
II)  Considering the high overhead of a large number of workers will join in the federated learning, efficient schemes for optimizing the number of workers are worth investigation in order to balance the learning performance and resource cost.
III)  It still remains to be an open issue on how to dynamically optimize the reputation threshold to minimize  negative effects from  malicious workers, e.g., by using advanced machine leaning methods.

\section*{Acknowledgement}
This work was supported in part by Singapore NRF National Satellite of Excellence, Design Science and Technology for Secure Critical Infrastructure NSoE DeST-SCI2019-0007, A*STAR-NTU-SUTD Joint Research Grant Call on Artificial Intelligence for the Future of Manufacturing RGANS1906, WASP/NTU M4082187 (4080), Singapore MOE Tier 1 2017-T1-002-007 RG122/17, MOE Tier 2 MOE2014-T2-2-015 ARC4/15, Singapore NRF2015-NRF-ISF001-2277, and Singapore EMA Energy Resilience NRF2017EWT-EP003-041. National Natural Science Foundation of China under Grant 61601336.

\section*{Biographies}
\small

{JIAWEN KANG} (kavinkang@ntu.edu.sg) received the M.S. degree from the Guangdong University of Technology, China, in 2015, and the Ph.D. degree at the same school in 2018. He is currently  a postdoc at Nanyang Technological University, Singapore.
His research interests mainly focus on blockchain, security and privacy protection in wireless communications and networking.

{ZEHUI XIONG} [S'17] (zxiong002@e.ntu.edu.sg) received his B.Eng. degree with the highest honors in Telecommunication Engineering from Huazhong University of Science and Technology, Wuhan, China, in 2016. He is currently working towards the Ph.D. degree in the School of Computer Science and Engineering, Nanyang Technological University, Singapore. He is a visiting student at Princeton University  in 2019. His research interests include network economics, wireless communications, blockchain, and deep reinforcement learning.

{DUSIT NIYATO}  [M'09, SM'15, F'17] (dniyato@ntu.edu.sg) is currently a professor in the School of Computer Science and Engineering, Nanyang Technological University. He received his B.Eng. from King Mongkut's Institute of Technology Ladkrabang, Thailand, in 1999 and his Ph.D. in electrical and computer engineering from the University of Manitoba, Canada, in 2008. His research interests are in the area of energy harvesting for wireless communication, the Internet of Things, and sensor networks.

{YUZE ZOU} (zouyuze@hust.edu.cn) received the B.E. degree in electronic information engineering  (EIE)  from Huazhong University of Science and Technology, Wuhan, China, in 2015, where he is currently pursuing the Ph.D. degree with the Department of EIE. His research interests include wireless power transfer, backscatter communications, and game theory and its applications in networked systems.

{YANG ZHANG} [M'11] (yangzhan2@whut.edu.cn)  received the B.Eng. degree from Beihang University, and the Ph.D. degree from Nanyang Technological University, Singapore. He is currently an associate professor at Wuhan University of Technology, China, and a research fellow in Nanyang Technological University, Singapore. His current research interests include market-oriented modeling for network resource allocations,  multiple objective optimization, and deep reinforcement learning.

{MOHSEN GUIZANI} [M'89, SM'99, F'09] (mguizani@ieee.org) received all of his degrees from Syracuse University, New York, in 1984, 1986, 1987, and 1990, respectively. He is currently with the Computer Science and Engineering Department, College of Engineering, Qatar University. He serves on the Editorial Boards of several international technical journals, and is the founder and Editor-in-Chief of Wireless Communications and Mobile Computing (Wiley). His research interests include wireless communications, mobile computing, computer networks, IoT, security, and smart grid.

%\bibliography{myreference}

\begin{thebibliography}{10}
	\bibitem{kang2018towards}
	J.~Kang~\emph{et~al.},  ``Towards secure blockchain-enabled internet of vehicles: Optimizing consensus management using reputation and contract theory,'' {\em IEEE Transactions on Vehicular	Technology}, vol. 68, no. 3, March 2019, pp. 2906-2920.
	
	\bibitem{cui2018survey}
	L.~Cui~\emph{et~al.}, ``A survey on application	of machine learning for internet of things,'' {\em International Journal of Machine Learning and Cybernetics}, 2018, pp.~1--19.
	
	\bibitem{dibconsortium}
	X.~Zhu~\emph{et~al.}, ``Blockchain-based privacy preserving deep learning,'' \textit{ International Conference on Information Security and Cryptology},  Springer, Cham, 2018, pp.~370-383.
	
	
	\bibitem{anh2018efficient}
	T.~T. Anh~\emph{et~al.},  ``Efficient	training management for mobile crowd-machine learning: A deep reinforcement	learning approach,'' in \textit{IEEE Wireless Communications Letters}, in press, 2019. DOI: 10.1109/LWC.2019.2917133
	
	\bibitem{shayan2018biscotti}
	M.~Shayan~\emph{et~al.},  ``Biscotti: A ledger for	private and secure peer-to-peer machine learning,'' 2018. [online] Available: \url{https://arxiv.org/abs/1811.09904}
	
	\bibitem{fung2018mitigating}
	C.~Fung~\emph{et~al.},  ``Mitigating sybils in federated learning poisoning,'' 2018. [online] Available: \url{ https://arxiv.org/abs/1808.04866}
	
	
	\bibitem{iotj2018}
	J.~Kang~\emph{et~al.}, 	``Incentive mechanism for reliable federated learning: A joint optimization approach to combining reputation and contract theory,'' {\em IEEE Internet of Things Journal}, in press, 2019. DOI: 10.1109/JIOT.2019.2940820
	
	\bibitem{liang2018towards}
	L.~Liang~\emph{et~al.}, ``Towards intelligent vehicular networks: A machine learning framework,'' \textit{IEEE Internet of Things Journal}, vol. 6, no. 1, 2019, pp. 124-135.
	
%    \bibitem{sg1}
%	 Q. Wang~\emph{et~al.}, ``Review of the false data injection attack against the cyber-physical power system,” in \textit{IET Cyber-Physical Systems: Theory \& Applications}, vol. 4, no. 2, pp. 101-107,	2019.
	 
	 \bibitem{sg2}
	  P. Zhuang~\emph{et~al.}, ``False Data Injection Attacks Against State Estimation in
	Multiphase and Unbalanced Smart Distribution Systems,” in \textit{IEEE Transactions on Smart Grid}, in press,
	2019. DOI: 10.1109/TSG.2019.2895306
	
%	\bibitem{nishio2018client}
%	T.~Nishio~\emph{et~al.}, ``Client selection for federated learning with	heterogeneous resources in mobile edge,''  \textit{IEEE International Conference on Communications (ICC)}, Shanghai, China, 2019, pp. 1-7.
	
	\bibitem{twoconsensus}
	J.~An~\emph{et~al.}, ``Crowdsensing	quality control and grading evaluation based on a two-consensus blockchain,''	{\em IEEE Internet of Things Journal }, vol. 6, no. 3, June 2019, pp. 4711-4718.
	
	\bibitem{2017quantifying}
	M.~Pouryazdan~\emph{et~al.},  ``Quantifying	user reputation scores, data trustworthiness, and user incentives in mobile crowdsensing,'' {\em IEEE Access}, vol.~5, 2017, pp.~1382--1397.
	
	\bibitem{xie2015incentive}
	H.~Xie~\emph{et~al.}, ``Incentive and reputation mechanisms for	online crowdsourcing systems,'' in {\em Quality of Service (IWQoS), 2015 IEEE 23rd International Symposium on}, IEEE, 2015, pp.~207--212.
	
	\bibitem{kim2018device}
	H.~Kim~\emph{et~al.},  ``Blockchained On-Device Federated Learning," in \textit{IEEE Communications	Letters}, in press, 2019. DOI: 10.1109/LCOMM.2019.2921755
	

	\bibitem{zhao2018federated}
	Y.~Zhao~\emph{et~al.}, ``Federated learning	with non-iid data,'' 2018. [online] Available: https://arxiv.org/abs/1806.00582 
	
	\bibitem{yang2018blockchain}
	Z.~Yang~\emph{et~al.}, ``Blockchain-based 	decentralized trust management in vehicular networks,'' {\em IEEE Internet of Things Journal},  vol. 6, no. 2, April 2019, pp. 1495-1505.
	
\end{thebibliography}
\end{document}